\documentclass[useAMS,usenatbib,usegraphicx]{mn2e}

\title[The radio lighthouse CU Vir]{The radio lighthouse CU Virginis:
\\ the spindown of a single main sequence star}
\author[C. Trigilio et al.]{
C. Trigilio$^{1,2}$\thanks{E-mail:email@address ctrigilio@oact.inaf.it},
P. Leto$^3$,
G. Umana$^1$,
C.S. Buemi$^1$,
F. Leone$^2$\\
$^{1}$INAF - Osservatorio Astrofisico di Catania, via S.Sofia 78, 95123 Catania, Italy\\
$^{2}$Dipartimento di Fisica ed Astronomia, Universit\`a di Catania, via S. Sofia 78, 95123 \\
$^{3}$INAF - Istituto di Radioastronomia, Sezione di Noto, CP 161, Noto (SR), Italy 
Catania, Italy}
\begin{document}

\date{Accepted  Received ; in original form}

\maketitle

\label{firstpage}

\begin{abstract}
The fast rotating star CU Virginis is a magnetic chemically peculiar star with an oblique 
dipolar magnetic field. The continuum radio emission has been interpreted as gyrosyncrotron emission 
arising from a thin magnetospheric layer. 
Previous radio observations at 1.4~GHz showed that a 100\% circular polarized and highly directive 
emission component overlaps to the continuum emission two times per rotation, when the magnetic axis 
lies in the plane of the sky.
This sort of radio lighthouse has been proposed to be due to cyclotron maser emission generated above the 
magnetic pole and propagating perpendicularly to the magnetic axis.
Observations carried out with the Australia Telescope Compact Array at 1.4 and 2.5~GHz one year after this
discovery show that this radio emission is still present, meaning that the phenomenon responsible for
this process is steady on a timescale of years. The emitted radiation spans at least 1 GHz, being 
observed from 1.4 to 2.5~GHz. On the light of recent results on the physics of the 
magnetosphere of this star, the possibility of plasma radiation is ruled out.
The characteristics of this radio lighthouse provides us a good marker of the rotation period, since the 
peaks are visible at particular rotational phases. After one year, they show a delay of about 15 minutes.
This is interpreted as a new abrupt spinning down of the star. Among several possibilities, a quick 
emptying of the equatorial magnetic belt after reaching the maximum density can account for the magnitude 
of the breaking.
The study of the coherent emission in stars like CU~Vir, as well as in pre main sequence stars,
can give important insight into the angular momentum evolution in young stars.
This is a promising field of investigation that high sensitivity radio interferometers such as SKA
can exploit.

\end{abstract}


\section{Introduction}
CU Virginis (=HD124224) is an A-type magnetic chemically peculiar star (MCP) with a
rotational period of 0.52 days, one of the shortest for this class of stars.
As observed in all MCP stars, the variability of the light curve is correlated 
with the spectroscopic variations \citep{deutsch, hardie}
and with the effective magnetic field \citep{borra80}.
All those characteristics can be explained in the framework of the oblique rotator model,
where the axis of the dipolar magnetic field is tilted with respect to the rotational 
one \citep{babcock} and abundance of elements is not homogeneously 
distributed over the stellar surface. 
The observed variabilities are thus consequence of the stellar rotation.

\begin{figure*} 
\includegraphics[width=17cm]{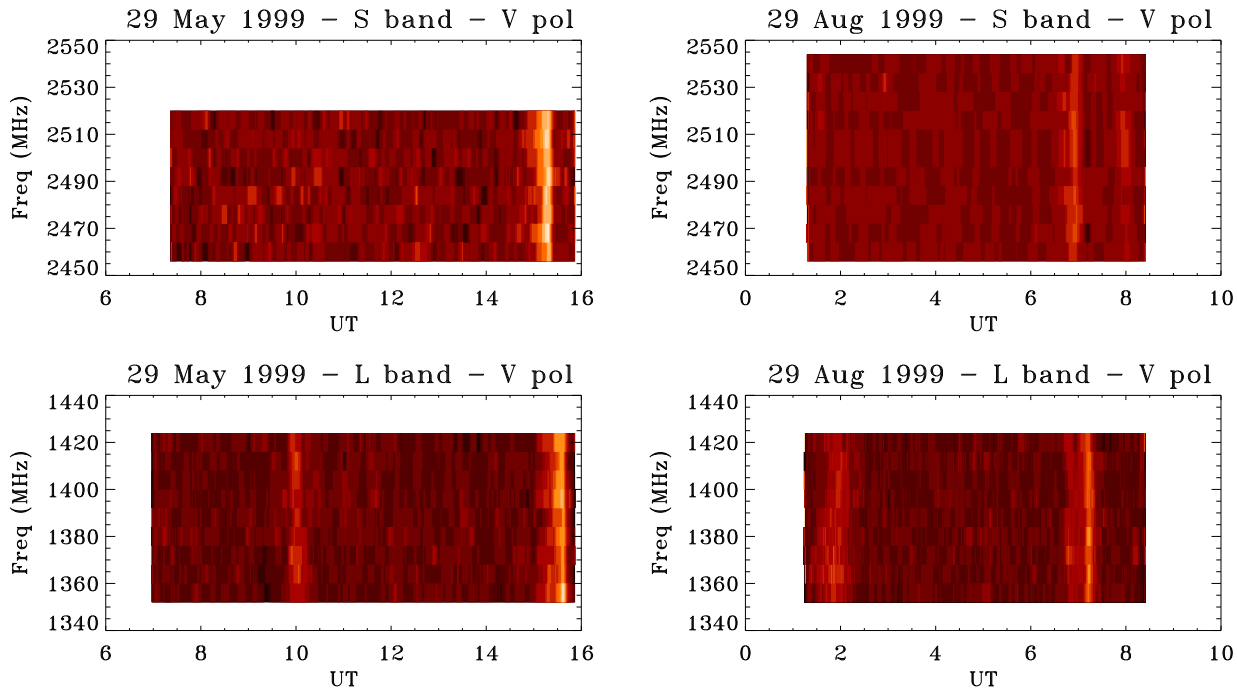}
      \caption{Dynamical spectra of CU~Vir during the two days
        of observations (left panels: May 29, 1999) at 2.5~GHz (upper panels)
        and 1.4~GHz (lover panels). Some spectral channels have been removed, reducing
        the bandpass. The spectral resolution is 8~MHz; the spectra have been smoothed
        in time with a window 2 minutes wide.
        Strong enhancements of the radio emission are evident at two
        phases at 1.4~GHz, while only one peak is visible at 2.5~GHz.
        }
  \label{spettri}
\end{figure*}

MCP stars are in general slow rotators in comparison with normal B and A-type main sequence stars.
This behaviour can be explained as the result of the action of a magnetic breaking.
But, at the present, only two MCP stars have been found to increase their rotational period; 
they are 56 Ari and CU Vir. While the former shows a continuous breaking down at the rate of few 
seconds per century \citep{adelman}, CU Vir has been subject of an abrupt increase of the rotational
period \citep{pyper}. From the analysis of photometric light curves
collected over 40 years, it seems that a change of period of about two seconds
occurred abruptly in 1984. The mechanism responsible for this event is still under
debate, and precise timing of the rotation is needed in order to detect any
further slowing down of the star.

Radio emission has been observed in CU Vir \citep{leo94}. The radio spectrum is quite flat
and extends up to mm wavelengths \citep{leo2004} with an high degree of circular
polarization \citep{leo96}. The variabilities of total intensity and polarization
are both correlated with the rotation of the star, suggesting we are in presence of 
gyrosyncrotron emission from an optically thick source. 
The anisotropic stellar wind inferred from spectral line
variations in MCP star \citep{shore}, associated with the magnetic field 
justifies the radio emission.

\citet{trig04} developed a three-dimensional model to explain the radio emission from MCP stars. 
The dipolar magnetic field interacts with the
stellar wind, that can freely escape from the polar regions outside the Alfv\'en
surface (outer magnetosphere) but remains trapped in the equatorial belt (inner magnetosphere). 
Ionized particles flowing out in the transition region between outer and inner magnetosphere,
called middle magnetosphere, 
stretch the magnetic field lines just outside the Alfv\'en radius and open the field
generating current sheets, where particles are accelerated up to relativistic energies.
They eventually propagate back toward the stellar magnetic poles following the field
lines and radiating for gyrosyncrotron process. As the magnetic field intensity increases
traveling to the star, they are reflected back outward by the magnetic mirroring effect
and are definitively lost from the magnetosphere.
This model, used to explain the observed fluxes and variability of MCP stars 
(HD37479 and HD37017) has been successfully applied to CU Vir by \citet{leto06}.
On the basis of multiwavelengths radio observations, important physical parameters of the stellar
magnetosphere, as the Alfv\'en radius ($12-17\,R_\ast$) and the mass loss (about $10^{-12}M_\odot \rmn{yr}^{-1}$) have been inferred.

A further observational evidence supporting the picture outlined above is the discovery
of coherent, highly directive, 100\% polarized radio emission at 1.4~GHz \citep{trig00}. 
The two peaks of radio emission have been observed at the rotational phases 
when the magnetic axis of the dipole is perpendicular
to the line of sight. The two peaks have been observed in three observing runs spanning 
10 days, indicating that the emission mechanism is persistent at least in timescales of
weeks. This has been interpreted as Electron Cyclotron Maser Emission from a population 
of electrons accelerated in the current sheets at the Alfv\'en point, that developed a
loss cone anisotropy after the mirroring and masing in a direction almost perpendicular to
the motion, so to the field lines, just above the magnetic pole.

In this paper we present radio observations of CU Vir at 1.4 and 2.5~GHz carried 
out with the aim to confirm the maser emission and to study its spectrum.
The directivity of the radiation is used to check the rotational period as
the beam point toward the Earth two times per rotation. 

\section{Observations and data reduction}
The observations have been carried out at the Australia Telescope Compact Array (ATCA)
\footnote{The Australia Telescope Compact Array is part of the Australia Telescope which is 
funded by the Commonwealth of Australia for operations as a National Facility managed 
by CSIRO} 
in two days, the first on May 29th 1999, the latter on August 29th 1999. 
We used all the 6 telescopes for both the observing runs.
On the first run the array was in configuration 6A, with a minimum and maximum
antenna spacing of 628 and 5939~m; on the latter run in configuration 6D, with
antenna spacing in the range from 367 to 5878~m.
The acquisition has been obtained with the 20-cm/13-cm feedhorns receivers, observing
simultaneously at two frequencies, namely L band, centered at 1384 MHz, and 
S band, centered at 2496 MHz, and two linear polarizations.
For both frequencies we used a bandwidth of 128 MHz,
split in 16 channels, with a spectral separation of 8 MHz. 
All the four correlator cross products of linear polarizations 
(XX, YY, XY, YX) for each couple of telescopes have been obtained with an integration
time of 30 seconds.
The target source CU Vir has been observed in cuts of 20 minutes,
preceded and followed by 4 minutes observations of the compact quasar
{1406-076}, used as phase calibrator.
The amplitude scale has been determined by observations on the radio galaxy
{PKS B1934-638}, with a flux density of 14.9 and 11.6 Jy at L and S band
respectively.

Data reduction and editing has been performed by using the MIRIAD package.
All the channels have been inspected and some removed because of interferences,
reducing in some case the total bandwidth.
The four linear polarizations cross products have been converted into the Stokes
parameters $I,Q,U,V$ by using the standard calibration tasks.

Maps of {CU Vir} have been obtained at 1.4 and 2.5~GHz in Stokes $I$ for the two
days of observations. The average flux densities are shown in Table~\ref{flux}. 
Errors have been evaluated as 
$\sigma=\sqrt{(\Delta Imm)^2+(\sigma_\mathrm{cal}F)^2}$,
where $\Delta Imm$ is the rms of the map and $\sigma_\mathrm{cal}F$ is the error associated
to the flux calibration, assumed 5\% of the measured flux density.

\begin{table}
   \caption[]{Stokes $I$ flux densities}
    \label{flux}
      \begin{tabular}[]{ccc}
      \hline
      Date	   & $F_\mathrm{1.4}\: (\sigma)$ & $F_\mathrm{2.5}\: (\sigma)$ \\
                   &       mJy        &        mJy       \\
      \hline
      29 May 1999  &  3.47  (0.30) &  3.39 (0.23) \\
      29 Aug 1999  &  2.24  (0.18) &  2.93 (0.18) \\
      \hline
\end{tabular}
\end{table}

\begin{figure} 
\resizebox{9cm}{!}{\includegraphics{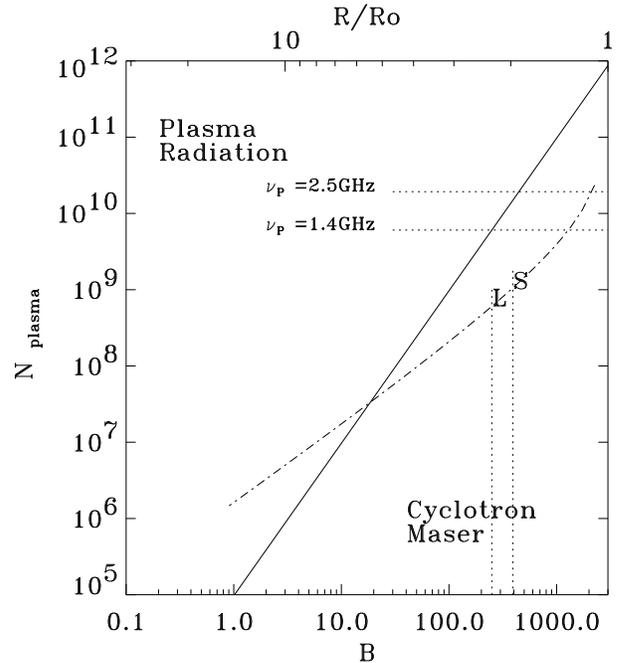}}
    \caption{Plane $B-N_\rmn{P}$: plasma radiation is possible only in the
    upper left part of the diagram; conversely, Cyclotron maser in the lower right
    part. The dashed line represents the plasma conditions along the field lines
    where the emitting particles flow. At the observed frequencies, only cyclotron
    maser is possible (points L and S).
    }
  \label{permesse}
\end{figure}

For each spectral channel $\nu$ and for each time $t$, the flux density $F(\nu,t)$ 
has been obtained by performing the DFT of the $N$ observed visibilities $V(\nu,t)$:
\[
F(\nu,t)=\frac{1}{N}\Sigma_{i=1}^N V_{i}(\nu,t)e^{-2\pi(u_{i}x_0+v_{i}y_0)}
\]
where $x_0$ and $y_0$ are the offsets of the target source from the
phase tracking center in RA and Dec respectively.

\subsection{Dynamical spectra}
The dynamical spectra in Stokes V for L and S band as a function of time are shown in
Fig.~\ref{spettri}. 
The circular polarization is almost zero except for limited periods of time.
In the 29 May observations, we note an enhancement of the Stokes V in the 2.5~GHz band
with a maximum at 15:10 UT, and that the flux density is almost constant inside this band
at a given time.
In the 1.4~GHz band two peaks are visible, the first at 10 UT, the latter at 15:26 UT, about 
16 minutes after the S band peak, and with a stronger intensity then the first one.
A similar behavior is evident during the 29 Aug observations, with one peak at 2.5~GHz
at 6:59 UT, and with two peaks at 1.4~GHz, at 1:54 UT and 7:13 UT, about 14 minutes after the 
2.5~GHz one.

\section{The cyclotron maser}
The first result of our observations is that the two enhancements of the radio emission discovered
in CU~Vir by \citet{trig00} are still present at 1.4~GHz after one year.
This means that the mechanism responsible for this coherent emission is persistent over 
a long time period. 
The radioemission has a broad band, since it is constant inside the observed band
and it is also visible at 2.5~GHz. 

In the following we will discuss in detail the behaviour of this component 
of radio emission and its implications.
\begin{figure*} 
\resizebox{17cm}{!}{\includegraphics{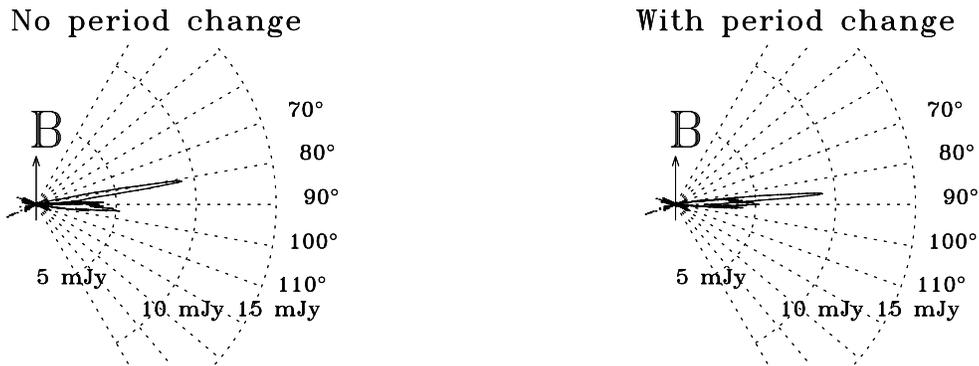}}
    \caption{Polar diagram of the radio emission at 1.4~GHz in the Aug 29 observations. 
    The magnetic axis of the dipole is vertical; 
    the emission is plotted as a function of the
    angle $\theta_\mathrm{M}$ formed by the line of sight and the
    axis B. 
    Left panel: $\theta_\mathrm{M}$ is computed by using the ephemeris given
    by \citet{pyper}. The direction of propagation of the radiation  
    is not the same for the two peaks.
    Right panel: assuming a delay of about 13 minutes, the direction of propagation
    of the two peaks coincide. In this case an abrupt change of period of about 1 second 
    is assumed.
    The behavior of the May 29 data is the same.}
  \label{beam}
\end{figure*}

\subsection{ECME vs Plasma radiation}
Two types of coherent radio emission are possible in a magnetized plasma; they are
Electron Cyclotron Maser Emission (ECME) and plasma radiation due to Langmuir waves.
In the ECME mechanism, electrons reflected by a magnetic mirror can develop
a pitch angle anisotropy. When energetic electrons propagate in a magnetic flux tube,
with an initial pitch angle $\psi_0$, they can penetrate inside the magnetosphere up 
to a point where $B=B_0/\sin \psi_0^2$, where $B_0$ is the magnetic field at the 
point of injection.
Electrons with a very small $\psi_0$ can propagate to the stellar photosphere,
collide with the local thermal plasma and can not be reflected back outward,
generating an hollow cone in the pitch angle distribution of the reflected particles.

\subsection{The directivity of the polarized emission}

\begin{figure} 
\resizebox{9cm}{!}{\includegraphics{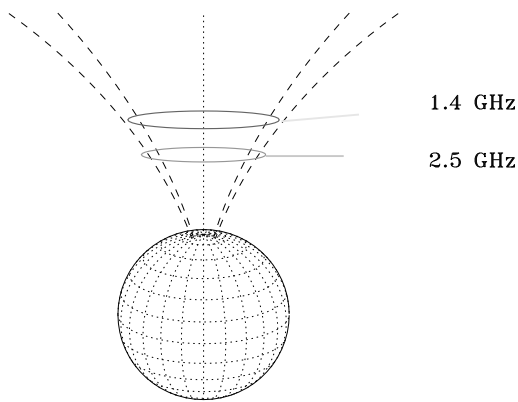}}
       \caption{
       Picture of the cyclotron maser emission: electrons accelerated in the current sheets
       out of the Alfv\'en radius propagate back to the star inside the thin layer 
       (the middle magnetosphere) between the two dashed lines; they are reflected back by magnetic
       mirroring, developing a loss cone anisotropy and emitting coherent radiation by cyclotron
       maser. Here the Alfv\'en radius is assumed to be 15 $R_\ast$, the radiation is emitted at
       harmonic number s=2, the 2.5~GHz radiation forms close to the star, while the 1.4~GHz at larger
       distance, where the magnetic field is weaker.
      }
  \label{sfera}
\end{figure}

\begin{figure*} 
\includegraphics[width=17cm]{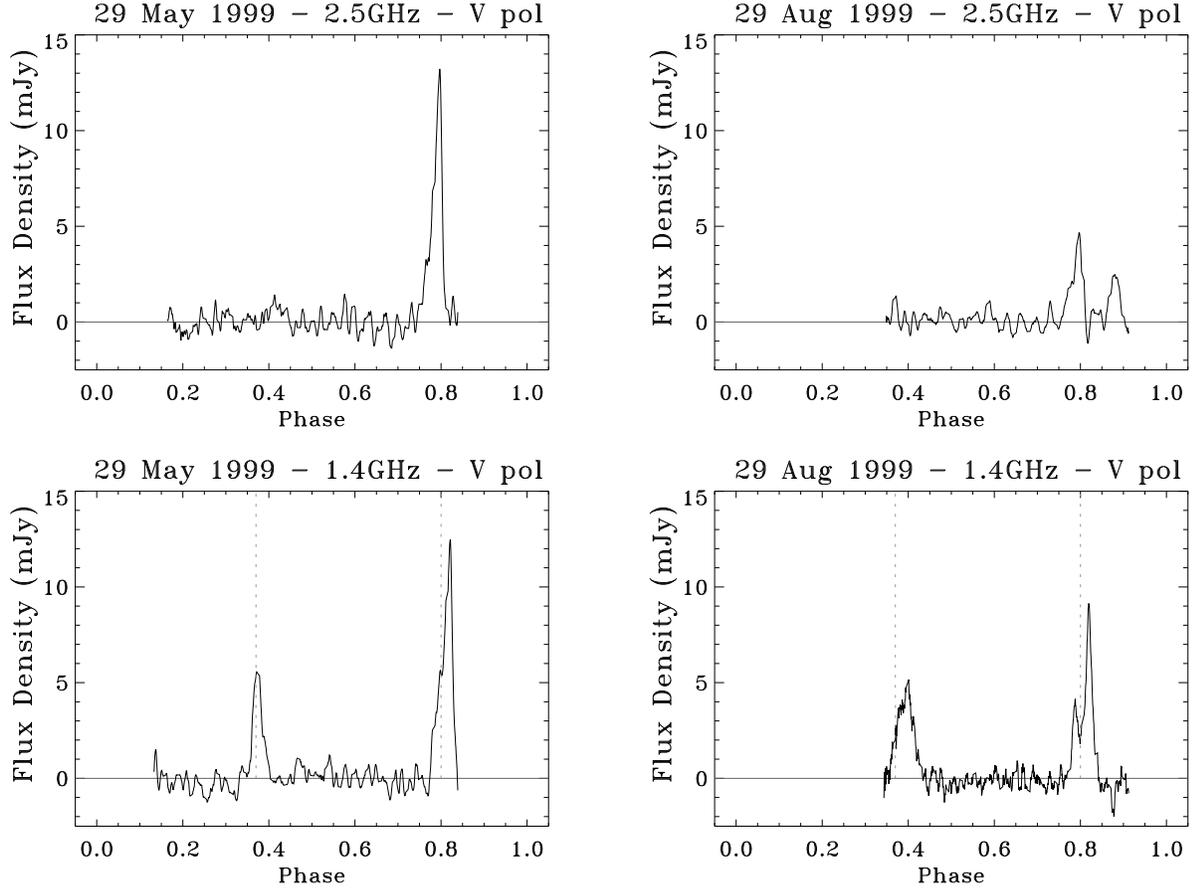}
       \caption{Flux density (Stokes V) at 2.5~GHz and 1.4~GHz as a function 
        of the rotational phase in the two days of observation. 
        The vertical dotted lines in the lower panels indicate the phases 
      of the peaks observed in 1998, showing the delay cumulated in one year
      because of a new increase of the rotational period.
      }
  \label{phases}
\end{figure*}

In this condition, if the plasma density is relatively small and the condition 
$\nu_\rmn{P}<<\nu_\rmn{B}$ can be satisfied
(where $\nu_\rmn{P}=9000\sqrt{N_\rmn{e}}\,$Hz is the plasma frequency, 
$N_\rmn{e}$ the plasma density number in $\rmn{cm}^{-3}$, and 
$\nu_\rmn{B}=2.8\times 10^{6}B\,$Hz is the gyrofrequency, with $B$ in gauss)
ECME can arise. 
The frequency of the maser emission is $\nu \ga s\;\nu_\rmn{B}$, where $s$ is 
the harmonic number ($s\leq 4$ in ECME). Since the first harmonic is generally suppressed 
by the gyromagnetic absorption of the local plasma, most of the emission is emitted at $s=2$.
The stimulated emission is directed almost
perpendicularly to the electron travel direction, which in turn is the
magnetic field line. Following \citet{melrose}, the radiation is confined in a hollow
cone of aperture $\theta$ with respect to the field line, with $\cos \theta=v/c$,
$v$ the average speed of the emitting electrons along the field line.
The beamwidth of this cone is $\Delta \theta\approx v/c$, and if the magnetic field
is constant in the region, $\Delta \nu/\nu \approx \cos^2\theta$.

In the hypothesis of the 3D model \citep{leto06} applied to CU Vir, 
the electrons accelerated in the current sheets travel through the middle magnetosphere.
Here the local plasma is the ionized wind, that increases moving inwards as the section
of the magnetic tube decreases ($N_\rmn{plasma}\propto r^{-3}/v_\rmn{wind}$). 
The corresponding plasma frequency in the inner magnetosphere varies as $r^{-3/2}$,
while the gyrofrequency varies like the magnetic field strength, i.e. as $r^{-3}$.
Therefore, moving inwards, the conditions for ECME are satisfied. 
The situation is shown in Fig.~\ref{permesse},
where plane $B$ vs $N_\rmn{plasma}$ identifies the status of a magnetoactive plasma
to be subject to plasma radiation or ECME. The diagonal line represents the loci where
the condition $\nu_\rmn{P}=\nu_\rmn{B}$ is valid. The shaded-dotted line represents 
the middle magnetosphere, i.e. the path of the electrons, from the Alfv\'en radius 
(lower left corner) to the stellar surface (upper right corner). 
The plasma density has been derived from \citet{leto06},
with a number density $N_\rmn{alf}=7\times10^5\;\rmn{cm}^{-3}$ at $R_\rmn{alf}=15\,R_\ast$.
The number density for plasma emission at 1.4 and 2.5 GHz is $N_\rmn{plasma}\approx 10^{10}\;\rmn{cm}^{-3}$
(horizontal dotted lines), but this condition is reached only very close to the stellar surface,
where the magnetic field strength is too high ($>1000$~Gauss) and only ECME is possible.

In conclusion, we rule out the possibility that the observed coherent emission is due to
plasma radiation. Only cyclotron maser is possible.

\citet{trig00} found that the polarized emission is highly directive, and it propagates
in direction  almost perpendicularly to the magnetic dipole axis. 
The observed peaks have been explained as due to 
the different inclination that the oblique dipole forms with the line of sight as the star rotates. 
At particular rotational phases, highly beamed radiation is emitted toward the Earth.
This happens two times per rotation, at two symmetric configurations.
The behavior is like a lighthouse, similar to a pulsar, even if the emission mechanism is different.

Since the flux density is constant inside each band (Fig.\ref{spettri}), it has been 
averaged in frequency in order to get a better S/N. Data have been folded using the 
ephemeris given by \citet{pyper}, used by \citet{trig00}, referred to light minimum, 
valid for $JD> 2\,446\,000$, nominally from 1984:
\[
HJD=2\,435\,178.6417+0^d.52070308 E
\]
The angle $\theta_\rmn{M}$ that the emitted radiation forms with respect to the 
magnetic axis is defined given the inclination $i$ of the rotational axis, the
obliquity $\beta$ of the magnetic axis
\[
\cos \theta_\rmn{M}=\sin \beta \sin i \cos(\phi-\phi_0)+\cos \beta \cos i,
\]
where $\beta=74\degr$, $i=43\degr$ and the phase $\phi_0=0.08$ corresponds to the delay
of the magnetic curve with respect to the light curve \citep{trig00}.
In Fig.~\ref{beam} the direction of the emitted radiation at 1.4~GHz for the 29 Aug observation
is shown as a function of the angle $\theta_\rmn{M}$. 
Because of the symmetry, the two beams should be emitted at the same angle. 
Looking at the left panel, however, it appears that the first beam is at
$\theta_\rmn{M}\approx 80\degr$, while the second at $\theta_\rmn{M}\approx 90\degr$.
This difference since the previous observations in 1998 can be explained if we assume that
the period has changed, and that the angle $\theta_\rmn{M}$ should be re-computed consequently.

The possibility of a change of period is very interesting since the discovery
of the abrupt increase of the rotational period of $\Delta P_1=2.18$~s
between 1983 and 1987 reported by \citet{pyper}. 
The use of the peaks of the coherent radiation as marker of the rotational period
is a powerful tool.

In order to get the new rotation period, we use the assumption that the two peaks must
be symmetric with respect to the dipole axis. In section~\ref{delay} we will compute the delay
accumulated in one year by evaluating the shift of the phase $\phi_0$.
In Fig.~\ref{beam}, right panel, data have been reported assuming the change of period
(see section~\ref{spindown}). 
Here $\theta_\rmn{M}\approx 85\degr$ at 1.4~GHz. In this geometry, 
at 2.5~GHz we get $\theta_\rmn{M}\approx 90\degr$.

\subsection{The bandwidth of the polarized emission}
From the dynamical spectra shown in Fig.~\ref{spettri} it is evident that the spectrum of the
radiation is constant inside each band. While the first peak is visible only at 1.4~GHz,
the second one is visible at the two bands, even if not exactly at the same times.

The theory of the ECME foresees a narrow bandwidth when the magnetic field is constant.
The existence of masing radiation in a broad range of frequency
means that it arises from the region above the pole where the magnetic field strength ranges 
from 450~G down to 250~G, i.e. at $R=1.9-2.8\,R_\ast$ when we assume 
$B_\mathrm{pole}=3000$~G and $s=2$. 
The scenario of the cyclotron maser emission is drawn in Fig.~\ref{sfera}.
Considering the angle between the magnetic field lines and the emitted radiation, we get
$\theta_\rmn{2.5 GHz}\approx 60\degr$ and $\theta_\rmn{1.4 GHz}\approx 50\degr$ for the 
radiation emitted at the two frequencies.

\subsection{The delay of the peaks at 1.4 GHz}
\label{delay}
The radio emission, averaged in frequency inside the observing bands, is shown in 
Fig.~\ref{phases} as a function of the rotational phase. This has been computed by using
the ephemeris of \citet{pyper}. 
\citet{trig00} found that in June 1998 the two main peaks were centered at phases 
$\phi_\mathrm{a}=0.37$ $\phi_\mathrm{b}=0.80$, with a phase difference between them of 0.43.
The present observations show that the phase difference between the peaks $a$ and $b$ 
is always the same; in fact we now measure 0.439 and 0.418. 
This indicates that the geometry of the magnetic field (dipole configuration, inclination $i$, 
obliquity $\beta$) is preserved.
In the lower panels of Fig.~\ref{phases}, where the 1.4~GHz curves are shown, the vertical lines mark the
phases of the main peaks observed in 1998. A delay is evident.

\begin{figure} 
\resizebox{9cm}{!}{\includegraphics{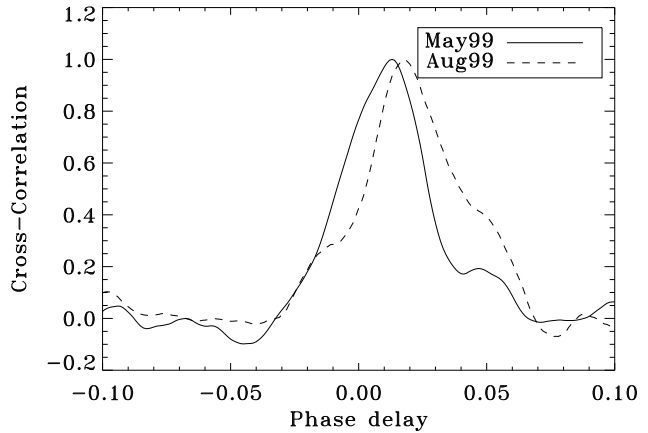}}
       \caption{
       Cross correlation between the two emission at 1.4 GHz (May99 and Aug99 data) and 
       the June 1998 data as a function of the rotational phase. The maximum of the functions 
       indicates the phase delay accumulated in one year because of the change of the 
       rotational period of the star.
      }
  \label{cross}
\end{figure}

In order to measure this delay, we search for the relative shift of the whole emission patters
as a function of the phase for the different observational sets. This can be done by computing the
cross correlations between the new curves with respect to the 1998 one. 
These are shown in Fig.~\ref{cross}.
Maxima and associate errors have been found by fitting the cross correlation functions 
at a level greather than 75\% of the amplitude with gaussians.
Results are listed in Table~\ref{shift}

\begin{table}
   \caption[]{Delay of the rotation markers at 1.4~GHz}
    \label{shift}
      \begin{tabular}[]{@{}lccc}
      \hline
      Date      &  Phase delay $d\phi$ &   $\sigma_{\phi}$ & Delay (min) \\
      \hline
      May 1999  &    0.0120            &    0.008          &  9 $\pm$ 6 \\
      Aug 1999  &    0.0180            &    0.008          & 13 $\pm$ 6 \\
      \hline
\end{tabular}
\end{table}

\section{The spin down of CU Vir}
\label{spindown}
The phase delay $d\phi$ of the markers of the rotation of CU Vir indicates
that the star is slowing down. 
New photometric data seem to indicate further slowdown \citep{pyper2004}.
But how does it happen? Continuously or suddenly?

In the case of a continuous breaking down, there should be a delay also in the
1998 observations with respect to the photometric observations reported by \citet{pyper}, 
that have been completed in 1997. In fact, the accuracy of the delay determination from
the radio observations allow to detect the slowing down in one year. But we have not
found any difference between the 1998 data and the photometric measurements of 1997, and
we can rule out the continuous breaking down.

\begin{figure} 
\resizebox{9cm}{!}{\includegraphics{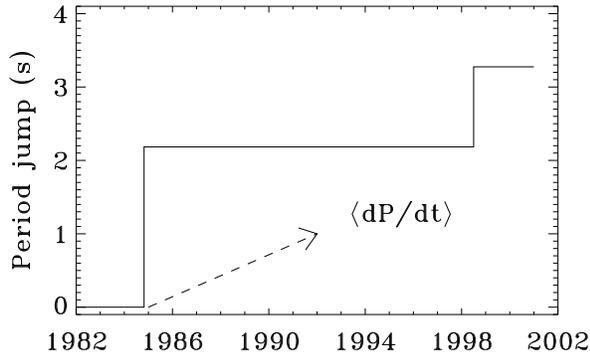}}
    \caption{Period jumps as a function of the time. 
    }
  \label{jumps}
\end{figure}

In the following we assume a sudden change of rotational period.
The average of the phase shifts during the two observations in 1999 is $d\phi=0.015$ and a delay $d\phi*P=0\fd0081\approx700\,\mathrm{s}$.
We don't know exactly when the change of period occurred; assuming that this occurred
in 1998, just after the previous observations, in $dt=1$ year (after $N_\mathrm{rot}=700$ 
rotations of the star) $P$ increased at least of 
\[
\Delta P_2=\frac{d\phi*P}{N_\mathrm{rot}} \ga 1 \,\mathrm{s}
\]
but if it occurred later, the change of period increases accordingly.
The change of rotation rate is thus:
\begin{equation}
\frac{\Delta P_2}{P} \ga 2\times 10^{-5}
\label{deltap}
\end{equation}
quite close to the previous period jump.
The period jumps are reported in Fig~\ref{jumps} as a function of the time.

The average spindown of the star can be evaluated by noting one period jump of 
$\Delta P_1=2.18$~s \citep{pyper} occurs in $\Delta t=15$ years. Since $\Delta P_1=2.18$~s;
this leads to an average breaking  given by
\begin{equation}
<\dot{P}>=\frac{\Delta P_1}{\Delta t}=5\times 10^{-9} \,\mathrm{s\,s^{-1}}
\end{equation}
We can compute the characteristic breaking rate
\begin{equation}
\frac{<\dot{P}>}{P}\approx 10^{-13} \mathrm{ s^{-1}}
\label{dp_su_p}
\end{equation}
corresponding to a characteristic breaking time $\tau\approx 2\times 10^5 \mathrm{years}$.
This behaviour seems to be confirmed by recent observations \citep{kellett} at a slightly 
different frequency.

Three possible explanations for the spin down mechanism acting in CU Vir exist:
\begin{enumerate}
\item there is a change of the moment of inertia due to a redistributions of the internal
mass of the star \citep{stepien2000};
\item the star loses angular momentum interacting with the external space via magnetic breaking;
\item violent emptying of the inner magnetosphere causes the loss of angular momentum,
as hypothesized by \citet{hg84}.
\end{enumerate}
The angular momentum $J=2\pi I/P$, where $I=k^2R_\ast^2M$ is the moment of inertia
with $kR_\ast$ radius of gyration ($k=0.3$ by \citet{stepien2000}) and $M$ mass of the star,
hence is 
\[
J\approx 1.7\times 10^{51} ~\mathrm{g\,cm^2\,s^{-1}}
\]
adopting $R_\ast=2.2 R_\odot$ and $M_\ast=3 M_\odot$ \citep{north}.
In the following we compute the change of rotational rate in the three hypothesis.

\subsection{Change of the internal structure}
If the star does not lose angular momentum, the change of
period is only due to the change of the moment of inertia 
\[
\Delta P/P=\Delta I/I=2\Delta k/k
\]
This possibility has been investigated by \citep{stepien}, who concluded that the magnetic field 
of CU Vir is too weak to alter the moment of inertia of the whole star considered as a rigid rotator.
However it should be possible a redistribution of the mass of the outer envelope of the star, 
where the magnetic energy dominates over the gravitational one.

\subsection{Steady spindown due to the massloss}
In the case that the star does not change its internal structure, $I$ is constant and
\[
\dot{J}/{J}=-\dot{P}/{P}.
\]
Now we consider the possibility that the spin down is due the wind flowing continuously 
from the magnetic polar caps. In this case however there should be no abrupt change of 
rotational period. The angular momentum losses is given by
\[
\dot{J}=\dot{M}R_\mathrm{alf}^2\omega \approx 5\times 10^{34} \mathrm{g\,cm^2\,s^{-2}}
\]
where $\dot{M}=10^{-12}M_\odot \mathrm{yr^{-1}}$ and $R_\mathrm{alf}=15\,R_\ast$ are the actual
mass loss and the Alfv\'en radius given by \citet{leto06}. We get 
\[
\dot{J}/J\approx 2\times 10^{-17} ~\mathrm{s^{-1}}
\]
that is 4 orders of magnitude lower that the value of 
equation (\ref{dp_su_p}). We can conclude that the steady loss of angular momentum due to the
stellar wind cannot explain the observed slowing down rotational period.

\subsection{Violent emptying of the inner magnetosphere}
We can instead assume that the material
accumulated in the inner magnetosphere \citep{leto06} reaches a maximum density,
than suddenly opens the field lines close to the Alfv\'en radius in a single event.
In this case
\[
\Delta J/J=-\Delta P/P.
\]
Following \citet{hg84}, the maximum density which can be kept by the magnetic field
in a rotating magnetosphere is given by 
$1/2 \rho_\mathrm{max}x^2\omega^2=B^2/8\pi $,
where $x$ and $B$ are the distance from the rotational axis and the magnetic field respectively.
In the inner magnetosphere, the magnetic field is given by 
$B=B_\mathrm{p}/(2r^{3})\sqrt{1+3\cos^2\theta}$, where $\theta$ is the magnetic colatitude 
and $B_\mathrm{p}$ the magnetic field strength at the pole.
We can get $\rho_\mathrm{max}(x,y)$ in all the points inside the inner magnetosphere, i.e. inside the 
magnetic line field reaching the Alfv\'en radius. 
The total mass that can be contained is given by
\[
\Delta M= 2\,\pi\,\int_{0}^{R_\rmn{Alf}}x \left[ \int_{0}^{h(x)} \rho(x,y)dy\right]  dx
\]
Adopting $R_\mathrm{alf}\approx 15 R\ast$ \citep{leto06}, in the magnetic equator, the maximum density
varies as $\rho_\mathrm{max}(x,0) \propto x^{-8}$, with a value at Alfv\'en radius of about 
$2\times 10^{-19}\rmn{g\,cm^{-3}}$ (corresponding to a number density of $10^5 \mathrm{cm^{-3}}$)
We get $\Delta M\approx 10^{25}$g.
However, as \citet{hg84} found that the density distribution in cool CP stars like CU~Vir deviates
from the Alfv\'en inside the magnetic loop of height about $5-8$ $R_\ast$, we get 
$\Delta M\approx 3\times 10^{24}$g.
If during this kind of violent event all this mass is released and flows through the Alfv\'en surface,
the corresponding jump of angular momentum is given by
\[
\Delta J=\Delta M R_\mathrm{alf}^2\omega\approx  2\times 10^{45} - 10^{46}~\mathrm{g~cm^2 s^{-1}}
\]
where the lower value refers to the case of magnetosphere not fully filled. In this hypothesis 
\[
\Delta J/J \approx 1.3\times 10^{-6} - 10^{-5}
\]
With all the uncertainty
of the case (in $R_\mathrm{alf}$, mass, magnetic field strength...) we can conclude that 
a violent release of material accumulated in the inner magnetosphere up to
the critical Alfv\'en density could account for the observed change of rotational period
(see eq.~\ref{deltap}).

\section{Conclusions}
The observations of CU Vir carried out with the ATCA at 1.4 and 2.5 GHz confirm the
presence of the coherent emission already reported by \citet{trig00} after one year from the
discovery, indicating that this is a steady phenomenon. 
The emitted radiation is visible only at rotational phases corresponding to the instant 
when the oblique axis of the magnetic dipole lies on the plane of the sky. 
This indicates the high directivity of this component of the radio emission.
All those characteristics, and the fact that it is 100\% right hand circularly polarized,
are in agreement with the process of electron cyclotron maser from electrons accelerated 
in the current sheets out of the Alfv\'en radius, propagating back to the stellar polar caps
and reflected outward by the converging magnetic field. The lack of reflected electrons at small
pitch angle, that fall in the stellar atmosphere, is the cause of the loss cone anisotropy 
that, in turn, generates this auroral radiation above the magnetic pole of the star.
The polarization properties, the fact that the radiation is only right hand polarized, 
means that this process in efficient only above the north magnetic pole. This is possible
as the magnetic field is not purely dipolar.
The possibility of plasma radiation is ruled out since the high magnetic field strength 
in the region where the radiation is generated.

Since the magnetosphere rotates obliquely around the rotational axis, the cyclotron maser
is visible only when it points toward the Earth, like a lighthouse. In this mode it is 
a good marker of the rotation of the star. The analysis done shows that the peaks are delayed
of about 15 minutes with respect to the observations carried out one year before, indicating 
a possible change of the rotational period of the star, of the order of 1 second, occurred 
in the period 1998--1999. A similar change of period \citep{pyper}, occurred around 1985, 
has been already reported. CU Vir is the unique single main sequence star with frequent 
abrupt spindown.

Different spinning down mechanisms are discussed: the possibility of a change of the
moment of inertia of the star, the continuous spindown due to the wind flowing from the
Alfv\'en surface and the violent emptying of the inner magnetosphere. 
In this latter hypothesis the material accumulated in the closed field lines of the equatorial
magnetic belt reaches a maximum density and opens the field lines in a violent event, releasing
an angular momentum which may account the observed breaking.

Cyclotron maser emission from stars provides important information on the magnetospheres,
as it has been observed in flare stars, in dMe and brown dwarfs. 
In the future, when high sensitivity radio interferometers such as SKA will allow
to discover more and more radio lighthouse of the same kind of CU Vir,
a new possibility to study with high precision the angular momentum evolution of young main 
sequence and pre main sequence stars will be opened.


\begin{thebibliography}{}
\bibitem[\protect\citeauthoryear{Adelman et al.}{2001}]{adelman}
        Adelman, S.J., Malanushenko, V., Ryabchikova, T.A.; Savanov, I., 
	2001, A\&A, 375, 982
\bibitem[\protect\citeauthoryear{Babcock}{1949}]{babcock} 
        Babcock, H.W. 1949, Observatory, 69, 191
\bibitem[\protect\citeauthoryear{Borra \& Landstreet}{1980}]{borra80} 
        Borra, E.F., Landstreet, J.D., 1980, ApJS, 42, 421
\bibitem[\protect\citeauthoryear{Deutsch}{1952}]{deutsch}
        Deutsch, A., 1952, ApJ, 116, 356
\bibitem[\protect\citeauthoryear{Hardie}{1958}]{hardie}
        Hardie, R., 1958, ApJ 127, 620
\bibitem[\protect\citeauthoryear{Havnes \& Goertz}{1984}]{hg84} 
        Havnes, O., Goertz, C.K. 1984, A\&A, 138, 421
\bibitem[\protect\citeauthoryear{Kellett et al.}{2007}]{kellett}
        Kellett, B.J., Graffagnino, V, Bingham, R., Muxlow, T.W.B., Gunn, A.G., 2007,
        astro-ph/0701214 
\bibitem[\protect\citeauthoryear{Leone et al.}{1994}]{leo94} 
        Leone, F., Trigilio, C., Umana, G. 1994, A\&A, 263, 908
\bibitem[\protect\citeauthoryear{Leone et al.}{2004}]{leo2004}
        Leone, F.; Trigilio, C.; Neri, R.; Umana, G., 2004, A\&A, 423, 1095
\bibitem[\protect\citeauthoryear{Leone et al.}{1996}]{leo96}
        Leone, F., Umana, G., Trigilio, C.,, 1996, A\&A, 310, 271
\bibitem[\protect\citeauthoryear{Leto et al.}{2006}]{leto06}
        Leto, P.; Trigilio, C.; Buemi, C. S.; Umana, G.; Leone, F. 2006
        A\&A, 458,831
\bibitem[\protect\citeauthoryear{Melrose \& Dulk}{1982}]{melrose}
        Melrose, D.B., Dulk, G.A., 1982, ApJ 207, 341
\bibitem[\protect\citeauthoryear{North}{1998}]{north}
        North, 1998, A\&A 334, 181
\bibitem[\protect\citeauthoryear{Pyper et al.}{1998}]{pyper} 
        Pyper, D.M., Ryabchikova, T., Malanushenko, V., Kuschnig, R.,
        Plachinda, S., Savanov, I. 1998, A\&A, 339, 822
\bibitem[\protect\citeauthoryear{Pyper \& Adelman}{2004}]{pyper2004}
        Pyper, D.M., Adelman, S.J., 2004, IAUS, 224, 307
\bibitem[\protect\citeauthoryear{Shore}{1987}]{shore} 
        Shore, S.N. 1987, AJ, 94, 731
\bibitem[\protect\citeauthoryear{St\c{e}pie\'n}{1998}]{stepien}
        St\c{e}pie\'n, K. 1998, A\&A 337, 756
\bibitem[\protect\citeauthoryear{St\c{e}pie\'n}{2000}]{stepien2000}
        St\c{e}pie\'n, K. 2000, A\&A 352, 227
\bibitem[\protect\citeauthoryear{Trigilio et al.}{2000}]{trig00}
        Trigilio, C., Leto, P., Leone, F., Umana, G., Buemi, C. 2000, A\&A, 362, 281 
\bibitem[\protect\citeauthoryear{Trigilio et al.}{2004}]{trig04}
        Trigilio, C., Leto, P., Umana, G., Leone, F., Buemi, C.S. 2004, A\&A, 418, 593 


\end{thebibliography}
\end{document}